\documentclass[preprint, 5p]{elsarticle}

\usepackage{amsmath,amssymb,bm}
\usepackage{lineno,hyperref}
\usepackage{multirow}
\usepackage{float}
\usepackage{booktabs}
\usepackage{array}

\begin{document}

\begin{frontmatter}

\title{Nuclear chiral rotation induced by superfluidity}

\author[PKU]{Y. P. Wang}
\author[PKU,Yukawa]{J. Meng\corref{cor1}}
\ead{mengj@pku.edu.cn}


\cortext[cor1]{Corresponding Author}
\address[PKU]{State Key Laboratory of Nuclear Physics and Technology, School of Physics, Peking University, Beijing 100871, China}
\address[Yukawa]{Yukawa Institute for Theoretical Physics, Kyoto University, Kyoto 606-8502, Japan}

\begin{abstract}
  The microscopic understanding on the influence of the pairing correlations or the superfluidity on the nuclear chiral rotation has been a longstanding and challenging problem. 
  Based on the three-dimensional cranking covariant density functional theory, a shell-model-like approach with exact particle number conservation is implemented to take into account the pairing correlations and applied for the chiral doublet bands built on the configuration $\pi h_{11/2}^2\otimes \nu h_{11/2}^{-1}$ in $^{135}$Nd. 
  The data available, including the $I-\omega$ relation, as well as the electromagnetic transition probabilities $B(M1)$ and $B(E2)$, are well reproduced. 
  It is found that the superfluidity can reduce the critical frequency and make the chiral rotation easier. 
  The mechanism is that the particle/hole alignments along the short/long axis are reduced by the pairing correlations, resulting in the enhanced preference of the collective rotation along the intermediate axis, and inducing the early appearance of the chiral rotation.
\end{abstract}

\end{frontmatter}


\section{Introduction}

The chirality is a subject of general interest in chemistry, biology, and physics. 
The concept of chirality in atomic nuclei was first proposed by Frauendorf and Meng in 1997 \cite{Frauendorf1997NPA}. 
They pointed out that, in the intrinsic frame of a rotating triaxial nucleus, the angular momentum vector may lie outside the three principal planes of the triaxial ellipsoidal density distribution. 
The short ($s$), intermediate ($i$), and long ($l$) principal axes of the triaxial nucleus form a screw with respect to the angular momentum vector, resulting in two systems with different intrinsic chirality, i.e., left- and right-handed systems. 
The broken chiral symmetry in the intrinsic frame would be restored in the laboratory frame, and this gives rise to the chiral doublet bands, i.e., a pair of nearly degenerate $\Delta I =1$ bands with the same parity.

In 2001, four pairs of nearly degenerate bands  were observed in the $N=75$ isotones $^{130}$Cs, $^{132}$La, $^{134}$Pr and $^{136}$Pm respectively, and were suggested as candidate chiral doublet bands \cite{Starosta2001PRL}.
In 2006, the electromagnetic transition ratios of the nearly degenerate bands in $^{134}$Pr were measured and their properties were in clear disagreement with the chiral interpretation \cite{Tonev2006PRL,Petrache2006PRL}.
This phenomenon was known as chiral conundrum.
In order to solve this problem, the adiabatic and configuration-fixed constrained triaxial relativistic mean field approaches were developed to search for nuclei with the chiral symmetry broken  \cite{MengJ2006PRC}.
A new phenomenon,  the multiple chiral doublets (M$\chi$D), i.e., more than one pair of chiral doublet bands in one single nucleus, was predicted for $^{106}$Rh \cite{MengJ2006PRC}.
In 2013, the first experimental evidence for M$\chi$D was reported in $^{133}$Ce \cite{Ayangeakaa2013PRL}.
Along this line, the lifetime measurements for the chiral doublet bands in $^{106}$Ag provided a resolution of the chiral conundrum, and the M$\chi$D with octupole correlations were first identified in $^{78}$Br \cite{Lieder2014PRL,LiuC2016PRL}.
As summarized in Ref. \cite{XiongBW2019ADNDT}, more than 59 chiral doublet bands in 47 nuclei (including 8 nuclei with M$\chi$D) have been reported in the $A\sim 80, 100, 130,$ and 190 mass regions.

The nuclear chirality has been extensively investigated with many theoretical approaches, including the triaxial particle rotor model (PRM) \cite{Frauendorf1997NPA,PengJ2003PRC,KoikeT2004PRL,ZhangSQ2007PRC,QiB2009PLB,ChenQB2018PLB,WangYY2019PLB,WangYP2020PRC},
the three-dimensional (3D) cranking model \cite{Madokoro2000PRC,Frauendorf1997NPA,Dimitrov2000PRL,Olbratowski2004PRL,ZhaoPW2017PLB,PengJ2020PLB}, the 3D cranking model with the random phase approximation \cite{Mukhopadhyay2007PRL,Almehed2011PRC}, the 3D cranking model with the collective Hamiltonian \cite{ChenQB2013PRC,ChenQB2016PRC,ChenQB2018PRC}, the interacting boson-fermion-fermion model  \cite{Brant2008PRC}, the generalized coherent state model \cite{Raduta2016JPG},
and the projected shell model \cite{Hara1995IJMPE,Bhat2014NPA,ChenFQ2017PRC,ChenFQ2018PLB,WangYK2019PRC}.
Among them, the 3D cranking relativistic \cite{ZhaoPW2017PLB,PengJ2020PLB} and  nonrelativistic \cite{Olbratowski2004PRL}  density functional theories can describe the nuclear chirality in a microscopic and self-consistent way, predict new chiral nuclei, and include important effects such as the core polarizations and nuclear currents \cite{MengJ2013FP,Olbratowski2004PRL,ZhaoPW2011PRL}.

The critical frequency (denoted as $\omega_{\text{crit}}$) is an important concept in the nuclear chiral rotation.
In Ref. \cite{Frauendorf1997NPA}, under the assumption of one $h_{11/2}$ particle and one $h_{11/2}$ hole coupled to a triaxial rotor, the angular momentum lying in the $sl-$plane at low rotational frequencies, will move out of the principal planes above $\omega_{\text{crit}}$, i.e., a transition from planar to aplanar rotation, in the tilted axis cranking (TAC) approximation. 
In the quantal PRM calculation, the chiral partner bands are nearly degenerate only within a limited spin range. 
The aplanar solution in the TAC approximation corresponds to the degenerate chiral doublets in the quantal PRM calculation. 
The occurrence of such an aplanar solution marks the breaking of the chiral symmetry in the intrinsic frame.  

The chiral critical frequency has been investigated by the 3D cranking models based on a Woods-Saxon potential combined with the Shell correction method \cite{Dimitrov2000PRL}, and relativistic \cite{ZhaoPW2017PLB,PengJ2020PLB} and  nonrelativistic \cite{Olbratowski2004PRL,Olbratowski2006PRC} density functional theories. 
However, in the microscopic and self-consistent density functional calculations, the pairing correlations are neglected.

In this work, based on the 3D cranking covariant density functional theory (CDFT), a shell-model-like approach (SLAP) with exact particle number conservation will be implemented to take into account the pairing correlations and applied for the chiral doublet bands. 
The longstanding and challenging problem about the microscopic understanding on the influence of the pairing correlations on the nuclear chiral rotation will be addressed.  

\section{Theoretical framework}

\subsection{3D cranking CDFT}

The 3D cranking CDFTs based on the meson-exchange or point-coupling interactions have been successfully applied to describe the chiral rotation, magnetic rotation, and linear chain structure of $\alpha$ clusters, etc \cite{Madokoro2000PRC,ZhaoPW2017PLB,ZhaoPW2015PRL}. 
The detailed formalism of the 3D cranking CDFT can be found in Refs. \cite{MengJ2013FP,ZhaoPW2017PLB}. Here, a brief introduction is presented.

The starting point of the CDFT is a universal density functional \cite{Ring1996PPNP,Vretenar2005PR,MengJ2006PPNP,Niksic2011PPNP,MengJ2016Relativistic}. 
In the 3D cranking CDFT, the functional is transformed into a body-fixed frame rotating with a constant angular velocity $\boldsymbol{\omega}$ around an arbitrary orientation in space. 
The corresponding Kohn-Sham equation for nucleons is a static Dirac equation,
  \begin{equation}\label{nucleonmotion}
     \hat{h}_0\psi_k
    =[ \boldsymbol{\alpha}\cdot
       (-i \boldsymbol{\nabla}-\boldsymbol{V})
       +\beta(m+S)+V
       -\boldsymbol{\omega}\cdot\boldsymbol{\hat{J}}]
     \psi_{k}
    =\varepsilon_{k}\psi_{k},
  \end{equation}
where $\psi_{k}(\boldsymbol{r})=\langle\boldsymbol{r}|\hat{b}_k^{\dagger}|0\rangle$, and $\hat{\boldsymbol{J}}$ is the nuclear total angular momentum. 
The scalar field $S$ and vector field $V^{\mu}$ are connected in a self-consistent way to the nucleon densities and currents. 
The nucleon densities and currents are obtained from the cranking single-particle eigenstates $\psi_{k}$. 
For example, the scalar density, $\rho_{S}(\boldsymbol{r})=\sum_{k>0}n_k \bar{\psi}_{k}(\boldsymbol{r}) \psi_{k}(\boldsymbol{r})$, in which $n_k$ is the occupation probability for state $\psi_k$, and the sum over $k > 0$ corresponds to the “no-sea approximation”.

The equation \eqref{nucleonmotion} can be solved by expanding the nucleon spinors in the three-dimensional Cartesian harmonic oscillator (3DHO) bases $\Phi_{\underline{\xi}}(\boldsymbol{r}, s)$ and $\Phi_{\overline{{\xi}}}(\boldsymbol{r}, s)$ \cite{PengJ2008PRC2},
  \begin{equation}\label{3DHO}
    \left\{ \begin{aligned}
            \Phi_{\underline{\xi}}(\boldsymbol{r}, s)
           =\phi_{n_{x}}(x) \phi_{n_{y}}(y) \phi_{n_{z}}(z) \frac{i^{n_{y}}}{\sqrt{2}}
            \left( \begin{array}{c}
                    1 \\
                    (-1)^{n_{x}+1}
                   \end{array}\right),&\\
            \text { \qquad\qquad\qquad\qquad\qquad~~ with } \underline{\xi}=\left|n_{x}, n_{y}, n_{z},+i\right\rangle, &\\
            \Phi_{\overline{\xi}}(\boldsymbol{r}, s)
          =\phi_{{n}_{x}}(x) \phi_{{n}_{y}}(y) \phi_{{n}_{z}}(z) \frac{(-i)^{{n}_{y}}}{\sqrt{2}}
           \left( \begin{array}{c}
                   (-1)^{{n}_{x}+1} \\
                   -1
                  \end{array}\right),&\\
            \text { \qquad\qquad\qquad\qquad\qquad~~ with } \overline{{\xi}}=\left|{n}_{x}, {n}_{y}, {n}_{z},-i\right\rangle,&
            \end{aligned}
    \right.
  \end{equation}
where $\Phi_{\underline{\xi}}(\boldsymbol{r}, s)=\langle \boldsymbol{r}, s|\hat{\beta}_{\underline{\xi}}^{\dagger}|0\rangle$, $\Phi_{\overline{\xi}}(\boldsymbol{r}, s)=\langle \boldsymbol{r}, s|\hat{\beta}_{\overline{\xi}}^{\dagger}|0\rangle$, and $\phi_{n_q}(q)$ is the normalized harmonic oscillator wavefunction for $q=x, y, z$. The wavefunctions $\Phi_{\underline{\xi}}(\boldsymbol{r}, s)$ and $\Phi_{\overline{{\xi}}}(\boldsymbol{r}, s)$ are the eigenstates of the simplex operator, $\hat{\mathcal{S}}=\hat{\mathcal{P}}\hat{\mathcal{R}}_1$, with the eigenvalues $+i$ and $-i$ respectively, where $\hat{\mathcal{P}}$ is the parity and $\hat{\mathcal{R}}_1$ is the signature operator, i.e., a rotation by $180^{\circ}$ around the $x$ axis. 
For the time-reversal operator, $\hat{\mathcal{T}}=i\sigma_y\hat{\mathcal{K}}$, with the complex conjugate operator $\hat{\mathcal{K}}$, one has $\hat{\mathcal{T}}\Phi_{\underline{\xi}}(\boldsymbol{r}, s)=\Phi_{\overline{{\xi}}}(\boldsymbol{r}, s)$ and $ \hat{\mathcal{T}}\Phi_{\overline{\xi}}(\boldsymbol{r}, s)=-\Phi_{\underline{\xi}}(\boldsymbol{r}, s)$. 
It means that $\Phi_{\underline{\xi}}$ and $\Phi_{\overline{\xi}}$ form a pair of time-reversal states up to a phase.

Diagonalizing the hamiltonian $\hat{h}_0$ in the 3DHO bases \eqref{3DHO}, the cranking single-particle eigenstates $\psi_k$ is obtained,  
  \begin{equation}\label{expansion}
    \begin{aligned}
       \psi_k
      &=\left( \begin{matrix}
                \sum_{\underline{\xi}} D^f_{\underline{\xi}k}\Phi_{\underline{\xi}}+\sum_{\overline{\xi}}D^f_{\overline{\xi}k}\Phi_{\overline{\xi}}\\
                \sum_{\underline{\tilde{\xi}}} D^g_{\underline{\tilde{\xi}}k}\Phi_{\underline{\tilde{\xi}}}+\sum_{\overline{\tilde{\xi}}}D^g_{\overline{\tilde{\xi}}k}\Phi_{\overline{\tilde{\xi}}}
              \end{matrix}\right),\\
       \text{~~or~~}\hat{b}_k^{\dagger}
      &=\left( \begin{matrix}
                \sum_{\underline{\xi}} D^f_{\underline{\xi}k}\hat{\beta}^{\dagger}_{\underline{\xi}}+\sum_{\overline{\xi}}D^f_{\overline{\xi}k}\hat{\beta}^{\dagger}_{\overline{\xi}}\\
                \sum_{\underline{\tilde{\xi}}} D^g_{\underline{\tilde{\xi}}k}\hat{\beta}^{\dagger}_{\underline{\tilde{\xi}}}+\sum_{\overline{\tilde{\xi}}}D^g_{\overline{\tilde{\xi}}k}\hat{\beta}^{\dagger}_{\overline{\tilde{\xi}}}
              \end{matrix}\right),\\
    \end{aligned}
  \end{equation}
where $D^{f}_{\underline{\xi}k}$ ($D^{g}_{\underline{\tilde{\xi}}k}$) and $D^{f}_{\overline{\xi}k}$ ($D^{g}_{\overline{\tilde{\xi}}k}$) are the expansion coefficients for the large (small) components.  

\subsection{Shell-model-like approach}

For open shell nuclei, the pairing correlations are taken into account by the SLAP \cite{ZengJY1983NPA,ZengJY1994PRC,MengJ2006FPC,ShiZ2018PRC}. 
The idea of the SLAP is to diagonalize the many-body Hamiltonian in a properly truncated many-particle configuration (MPC) space with exact particle number. 
The SLAP has many advantages, such as the conservation of the particle number, the strict treatment of the Pauli blocking effects, and the unified description for both the ground and excited states.

The cranking many-body Hamiltonian reads
  \begin{equation}\label{manybody-H}
    \hat{H}=\hat{H}_{\text{0}}+\hat{H}_{\text{pair}}.
  \end{equation}

The one-body Hamiltonian, $\hat{H}_{\text{0}}=\sum \hat{h}_0$, with $\hat{h}_0$ given in Eq. \eqref{nucleonmotion}, is diagonal in the cranking single-particle bases. 

For the two-body Hamiltonian, the long-range particle-hole correlations are already included in $\hat{H}_0$. 
For the short-range particle-particle correlations, a monopole pairing force with constant strength $G$ is used,
  \begin{equation}
    \hat{H}_{\text{pair}}=-G\sum_{\underline{\xi},\underline{\eta}>0}{\hat{\beta}_{\underline{\xi}}^{\dagger}
                          \hat{\beta}_{\bar{\xi}}^{\dagger}
                          \hat{{\beta}}_{\bar{\eta}}\hat{\beta}_{\underline{\eta}}},
  \end{equation}
where $\hat{\beta}^{\dagger}_{\underline{\xi}(\overline{\xi})}$ and $\hat{\beta}_{\underline{\eta}(\overline{\eta})}$ are the creation and annihilation operators for the 3DHO bases in Eq. \eqref{3DHO}.

From the expansions in Eq. \eqref{expansion}, $\hat{H}_{\text{pair}}$ can be rewritten in the cranking single-particle bases as,
  \begin{equation}
    \begin{aligned}
       \hat{H}_{\text {pair }}
      =-G \sum_{k_{1} k_{2} k_{3} k_{4}} &\left( \sum_{{\underline{\xi}},{\underline{\eta}}>0} D_{\underline{\xi}{k_1}}^{f*} D_{\overline{\xi}{k_2}}^{f*} 
                                                 D^{f}_{\overline{\eta}{k_4}} D^f_{\underline{\eta}k_{3}}\right.\\
                                                 &\left.+\sum_{{\underline{\xi}},{\tilde{\underline{\eta}}}>0} D_{\underline{\xi}{k_1}}^{f*} D_{\overline{\xi}{k_2}}^{f*} D^{g}_{\overline{\tilde{\eta}}{k_4}} D^g_{\underline{\tilde{\eta}}k_{3}}\right.\\
                                                 &\left.+\sum_{{\tilde{\underline{\xi}}},{\underline{\eta}}>0} D_{\underline{\tilde{\xi}}{k_1}}^{g*} D_{\overline{\tilde{\xi}}{k_2}}^{g*} D^{f}_{\overline{\eta}{k_4}} D^f_{\underline{\eta}k_{3}}\right.\\
                                                 &\left.+\sum_{{\tilde{\underline{\xi}}},{\tilde{\underline{\eta}}}>0} D_{\underline{\tilde{\xi}}{k_1}}^{g*} D_{\overline{\tilde{\xi}}{k_2}}^{g*} D^{g}_{\overline{\tilde{\eta}}{k_4}} D^g_{\underline{\tilde{\eta}}k_{3}}\right) \\
                                                 &{\hat{b}}_{k_{1}}^{\dagger} {\hat{b}}_{k_{2}}^{\dagger} {\hat{b}}_{k_{4}} {\hat{b}}_{k_{3}}.\\
    \end{aligned}
  \end{equation}

The MPC for the $A$-particle system in the cranking single-particle bases can be constructed as,
  \begin{equation}
     |i\rangle \equiv\left|k_{1} k_{2} \cdots k_{A}\right\rangle
    =\hat{b}_{k_{1}}^{\dagger} \hat{b}_{k_{{2}}}^{\dagger} \cdots \hat{b}_{k_{A}}^{\dagger}|0\rangle.
  \end{equation}

Diagonalizing the cranking many-body Hamiltonian \eqref{manybody-H} in the MPC space, the eigenstates are obtained,
  \begin{equation}
    \Psi=\sum_i C_i|i\rangle, 
  \end{equation}
with the expanding coefficients $C_i$.

From the expanding coefficients $C_i$, the occupation probability $n_k$ for the cranking single-particle state $\psi_k$ is,
  \begin{equation}
    n_{k}=\sum_i|C_{i}|^2P_{i}^{k}, 
    ~~P_{i}^{k}=\left\{ \begin{aligned}
                          &1, \text{~~} \psi_k \text{ is occupied in MPC } |i\rangle,\\
                          &0, \text{~~otherwise}. 
                        \end{aligned}\right.
  \end{equation}
The occupation probabilities are used in the calculations for the nucleon densities and currents, which will be interated back into Eq. \eqref{nucleonmotion} to achieve self-consistency \cite{MengJ2006FPC}.

From the nuclear eigenstates $\Psi$, the physical observables, including the angular momentum, the quadrupole moments and magnetic moments, and the electromagnetic transition probabilities $B(M1)$ and $B(E2)$, can be calculated.  

The expectation values of the angular momentum components are given as,
  \begin{equation}
    J_q = \langle\Psi|\hat{J}_q|\Psi\rangle,\quad q = x, y, z.
  \end{equation}
From the semiclassical relation, $\langle \hat{\boldsymbol{J}}\rangle^2=I(I+1)$, the quantized angular momentum $I$ is obtained.

The quadrupole moments $Q_{20}$ and $Q_{22}$ are,
  \begin{equation}
  \label{Q20}
    Q_{20}=\sqrt{\frac{5}{16\pi}}\langle 3z^2-r^2\rangle,
    \quad Q_{22}=\sqrt{\frac{15}{32\pi}}\langle x^2-y^2\rangle.
  \end{equation}

The nuclear magnetic moment in units of the nuclear magneton is given by, 
  \begin{equation}\label{mu}
    \begin{aligned}
       {\boldsymbol{\mu}}
      &=\sum_{k>0}n_k \int d^{3} r\left[ \frac{m c^{2}}{\hbar c} q \psi_{k}^{\dagger} \boldsymbol{r} \times \boldsymbol{\alpha} \psi_{k}
                                         +\kappa \psi_{k}^{\dagger} \beta \boldsymbol{\Sigma} \psi_{k}\right]\\
      &=\sum_{k>0}n_k \int d^{3} r\left[ \frac{qm}{m^*}\bar{\psi}_k(\boldsymbol{L}+\boldsymbol{\Sigma})\psi_k
                                         +\kappa \psi_{k}^{\dagger} \beta \boldsymbol{\Sigma} \psi_{k}\right],\\
    \end{aligned}
  \end{equation}
where the charge $q$ is 1 for protons and 0 for neutrons in units of $e$, $\kappa$ is the nucleon anomalous gyromagnetic factor, $\kappa_p=1.793$, $\kappa_n=-1.913$, $m^*$ denotes the Dirac effective mass, and $\boldsymbol{L}$ and $\boldsymbol{\Sigma}$ are respectively the orbital angular momentum and spin. 
In the usual CDFT, the effective mass $m^*\approx 0.58m$ \cite{ZhaoPW2010PRC}, because the Fock contributions are not taken into account. 
In the following calculations,  the ratio $m/m^*$ in Eq. \eqref{mu} is taken as 1, as Refs. \cite{WangYK2017PRC,ZhaoPW2017PLB}.

The transition probabilities $B(M1)$ and $B(E2)$ are calculated in the semiclassical approximation,
  \begin{equation}
    \begin{aligned}
      B(M 1)=\frac{3}{8 \pi}
             &\left\{ \left[ -\mu_{z} \sin \theta+\cos \theta\left(\mu_{x} \cos \varphi
                             +\mu_{y} \sin \varphi\right)\right]^{2}\right.\\
             &\left.+\left(\mu_{y} \cos \varphi-\mu_{x} \sin \varphi\right)^{2}\right\}, \\
    \end{aligned}
  \end{equation}
  \begin{equation}
    \begin{aligned}
      {B(E 2)}=&{\frac{3}{8}}\left[ Q_{20}^{p} \sin ^{2} \theta+\sqrt{\frac{2}{3}} Q_{22}^{p}
                                    \left(1+\cos ^{2} \theta\right) \cos 2 \varphi\right]^{2}\\
               &+\left(Q_{22}^{p} \cos \theta \sin 2 \varphi\right)^{2},\\
    \end{aligned}
  \end{equation}
where $Q_{20}^p$ and $Q_{22}^p$ are the quadrupole moments of protons, and $\theta$ and $\varphi$ are the orientation angles of the total angular momentum $\boldsymbol{J}$ in the intrinsic frame.
\section{Numerical Details}

In $^{135}$Nd, the first chiral doublets in odd-A nuclei were reported \cite{ZhuS2003PRL}, which was further affirmed by the lifetime measurements \cite{Mukhopadhyay2007PRL}. 
With a new pair of chiral doublet bands discovered in 2019 \cite{LvBF2019PRC}, $^{135}$Nd is supposed to be a new candidate with M$\chi$D. 
Theoretically, the chiral doublet bands in $^{135}$Nd have been investigated by different kinds of models, including the PRM \cite{QiB2009PLB}, the algebraic interacting boson-fermion model \cite{Brant2009PRC}, the TAC relativistic Hartree-Bogoliubov (RHB) \cite{ZhaoPW2015PRC}, the 3D cranking CDFT \cite{PengJ2020PLB}, and the time-dependent CDFT \cite{RenZX2022PRC}. 

Here, the 3D cranking CDFT-SLAP is applied for the chiral doublet bands built on the configuration $\pi h_{11/2}^2\otimes \nu h_{11/2}^{-1}$ in $^{135}$Nd. 
The point-coupling density functional PC-PK1 \cite{ZhaoPW2010PRC} is adopted in the particle-hole channel, and the monopole pairing interaction is adopted in the particle-particle channel. The number of major shells for the 3DHO bases is 10. 
The dimensions of the MPC space are 1000 for both the neutrons and protons. 
The effective neutron and proton pairing strengths $(\text{G}_\text{n}, \text{G}_\text{p})$ are ($0.55$MeV, $0.60$MeV), determined by reproducing the experimental odd-even mass differences around $^{135}$Nd. 
A larger MPC space with the renormalized pairing strengths according to the odd-even mass differences gives essentially the same results, indicating the convergence of the MPC space.

In the 3D cranking CDFT-SLAP, the quadrupole moments $Q_{2-2}$ and $Q_{2\pm 1}$ are constrained using the augmented Lagrangian method to define the intrinsic reference frame. The validity of the Kerman-Onishi \cite{Kerman1981NPA} (KO) conditions on the Lagrangian multipliers have been checked similarly as in Refs. \cite{ShiY2013PRC,ShiY2012PRL}. Under the convention for quadrupole moments in Eq. \eqref{Q20}, the KO conditions read,
\begin{equation}
\begin{split}
&L_{2-1}=-\frac{\omega_yJ_z-\omega_zJ_y}{\frac{1}{2}\sqrt{\frac{16\pi}{5}}{Q}_{20}+\frac{1}{2}\sqrt{\frac{32\pi}{15}}{Q}_{22}}\equiv L_{2-1}^{\prime}\\
&L_{21}=\frac{\omega_zJ_x-\omega_xJ_z}{\frac{1}{2}\sqrt{\frac{16\pi}{5}}{Q}_{20}-\frac{1}{2}\sqrt{\frac{32\pi}{15}}{Q}_{22}}\equiv L_{21}^{\prime}\\
&L_{2-2}=\frac{\omega_xJ_y-\omega_yJ_x}{\sqrt{\frac{32\pi}{15}}{Q}_{22}}\equiv L_{2-2}^{\prime}\\
\end{split}
\end{equation}
where $L_{2-1}, L_{21}$ and $L_{2-2}$ are the Lagrangian multipliers for $Q_{2-1}, Q_{21}$ and $Q_{2-2}$ respectively. Table \ref{table} lists the angular momentum components and the quadrupole moments, as well as the ratios $L_{2-1}/L_{2-1}^{\prime}$, $L_{21}/L_{21}^{\prime}$ and $L_{2-2}/L_{2-2}^{\prime}$ at $\hbar\omega = 0.4$MeV for the 3-qp band of $^{135}$Nd at the tilting angle $(\theta_{\omega}=30^{\circ}, \varphi_{\omega}=40^{\circ})$ with and without pairing. It can be seen that the KO conditions are satisfied with high precision with the values
$L_{2-1}/L_{2-1}^{\prime}$, $L_{21}/L_{21}^{\prime}$ and $L_{2-2}/L_{2-2}^{\prime}$ in the interval $[0.997, 1.002]$.
 \begin{table*}[h!t]
 \renewcommand\arraystretch{1.5}
 \label{table}
    \centering  
    \addtolength{\tabcolsep}{3pt}  
    \begin{tabular}{c c c c c c c}
        \hline\hline 
        & $( J_x, J_y, J_z)$ & $Q_{20}$ & $Q_{22}$     & $L_{2-1}/L_{2-1}^{\prime}$ & $L_{21}/L_{21}^{\prime}$ & $L_{2-2}/L_{2-2}^{\prime}$ \\
        \hline
        No Pairing           & (12.103, 5.705, 11.916) & 270.456 & -71.702  & 1.000 & 0.997 & 1.002 \\
         \hline
        Pairing              & (12.023, 5.900, 11.778) & 266.752 & -69.219  & 0.998 & 0.997 & 1.002 \\
        \hline \hline \\[-0.1cm]
    \end{tabular}
    \caption{ The angular momentum components and the quadrupole moments, as well as 
              the ratios $L_{2-1}/L_{2-1}^{\prime}$, $L_{21}/L_{21}^{\prime}$ and $L_{2-2}/L_{2-2}^{\prime}$ at $\hbar\omega = 0.4$MeV for the 3-qp band of $^{135}$Nd at the tilting angle $(\theta_{\omega}=30^{\circ}, \varphi_{\omega}=40^{\circ})$ with and without pairing.}
\end{table*}

\section{Results and discussions}\label{Sec4}

\begin{figure*}[ht]
  \centering
  \includegraphics[width=0.9\linewidth]{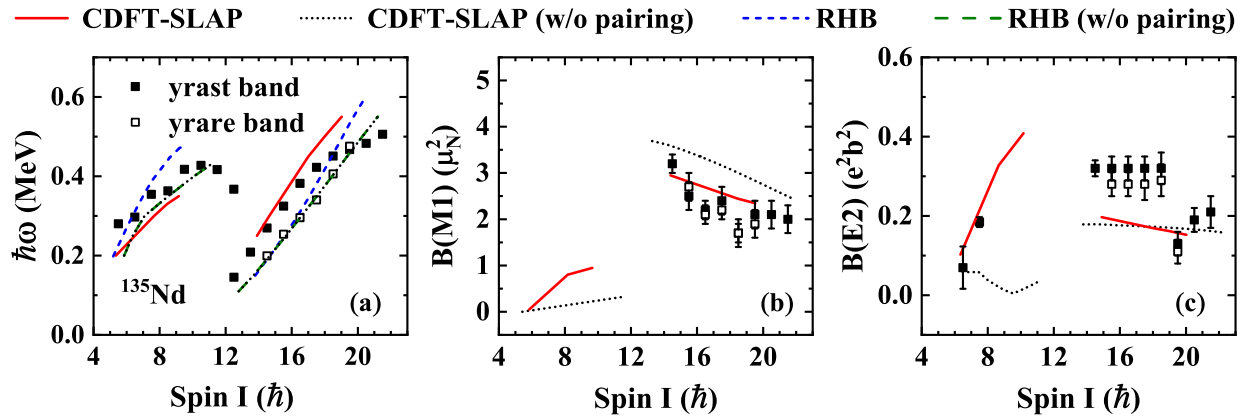}
  \caption{Rotational frequency (left), $B(M1)$ (middle) 
           and $B(E2)$ (right) values as functions of the 
           angular momentum in the 3D cranking CDFT-SLAP 
           calculations with and without pairing in comparison 
           with the data 
           (solid dots) \cite{ZhuS2003PRL,Mukhopadhyay2007PRL} 
           and the corresponding TAC-RHB 
           results \cite{ZhaoPW2015PRC}.}
  \label{fig1}
\end{figure*}

The angular momenta, and the electromagnetic transition probabilities $B(M1)$ and $B(E2)$ for the yrast band in $^{135}$Nd, calculated with the 3D cranking CDFT-SLAP with and without pairing, are shown in Fig. \ref{fig1}, in comparison with the data available~\cite{ZhuS2003PRL,Mukhopadhyay2007PRL}, and the corresponding TAC-RHB results in Ref.~\cite{ZhaoPW2015PRC}. 
The ground band of $^{135}$Nd is built on the one-quasiparticle (1-qp) configuration $\nu h_{11/2}^{-1}$ \cite{Beck1987PRL}. 
Above $I=25/2\hbar$, the band with the 3-qp configuration $\pi h_{11/2}^2\otimes \nu h_{11/2}^{-1}$ becomes yrast. 
It can be seen that, switching off the pairing, the results of the 3D cranking CDFT-SLAP and the TAC-RHB are the same, which confirms the correct implementation for the 3D cranking CDFT-SLAP. 
After the pairing correlations are included, better agreements with the experimental $I-\omega$ relations, and the $B(M1)$ and $B(E2)$ transitions, are achieved for both the 1-qp and the 3-qp band.

For the $I-\omega$ relations in Fig. \ref{fig1} (a), the 3D cranking CDFT-SLAP provides interesting physics for the 1-qp and 3-qp bands. 
For the 1-qp band, the angular momenta in the 3D cranking CDFT-SLAP are slightly larger than those in the TAC-RHB with and without pairing for given rotational frequency. 
This is due to the many-body correlations taken into account in the 3D cranking CDFT-SLAP. 
For the 3-qp band, the angular momenta in the 3D cranking CDFT-SLAP are reduced more significantly by pairing than those in the TAC-RHB for given rotational frequency, and a better description for the data is achieved. 
This is because the 3D cranking CDFT-SLAP correctly treats the pairing correlations with exact particle number conservation and avoid the proton pairing collapse occurring in the TAC-RHB.

For the $B(M1)$ transitions in Fig. \ref{fig1} (b), the results calculated by the 3D cranking CDFT-SLAP with and without pairing are compared with the data \cite{Mukhopadhyay2007PRL}. 
For the 1-qp band, the $B(M1)$ are enhanced due to the configuration mixing in the 3D cranking CDFT-SLAP, in comparison with the case without pairing. 
For the 3-qp band, the $B(M1)$ are suppressed in the 3D cranking CDFT-SLAP, because the pairing correlations reduce the transverse magnetic moment by merging the orientations of the proton and neutron angular momenta, similarly as Ref. \cite{ZhaoPW2015PRC}.

For the $B(E2)$ transitions in Fig. \ref{fig1} (c), the results calculated by the 3D cranking CDFT-SLAP with and without pairing are compared with the data \cite{Mukhopadhyay2007PRL}. 
For the 1-qp band, the $B(E2)$ are enhanced in the 3D cranking CDFT-SLAP and in better agreement with the data, because the  pairing correlations drive the angular momentum away from the $l$ axis, similarly as Ref. \cite{ZhaoPW2015PRC}. 
For the 3-qp band, the $B(E2)$ are insensitive to the pairing correlations, because the effects of the angular momentum moving away from the $l$ axis are compensated by the change of the quadrupole moments.

In Fig. \ref{fig1}, the data for the $I-\omega$ relation, and the $B(M1)$ and $B(E2)$ transition probabilities for the partner band of the 3-qp band are also shown. 
This pair of partner bands have similar $I-\omega$ relations and electromagnetic transitions, in agreement with the characteristics of chiral doublet bands. 
A kink around $I=37/2\hbar$ in the experimental $I-\omega$ relation appears in the yrast band. 
This kink is believed to be a sign of a transition from planar to aplanar rotation in the intrinsic frame \cite{Frauendorf1997NPA}. 
As the 3D cranking CDFT-SLAP allows for an arbitrary orientation of the angular momentum vector in the intrinsic frame, it provides a microscopic and self-consistent study of the transition from planar to aplanar rotation with pairing correlations. 

\begin{figure*}[!htbp]
  \centering
  \includegraphics[width=0.8\linewidth]{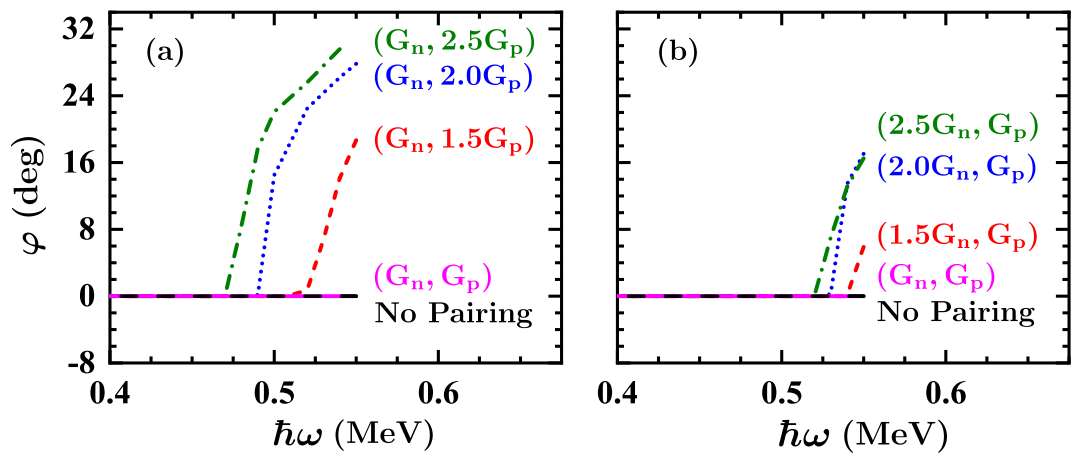}
  \caption{Evolution of the azimuth angle $\varphi$ for the total angular momentum as functions of the rotational frequency $\omega$ 
           in the cases without pairing or with different pairing strengths for protons (left panel) and neutrons (right panel).}
  \label{fig2}
\end{figure*}

The possible aplanar rotation can be examined from azimuth angle $\varphi$ of the total angular momentum $\boldsymbol{J}$. 
In Fig. \ref{fig2}, the evolution of $\varphi$ as functions of the rotational frequency are illustrated for different pairing strengths. 
Without pairing, the angle $\varphi$ is always zero with the rotational frequency. With the pairing strengths $(\text{G}_\text{n},\text{G}_\text{p})$, $\varphi$ still remains to be zero up to $\hbar\omega= 0.55$MeV. 
Beyond $\hbar\omega=0.55$MeV, it's difficult to achieve convergent results for the configuration $\pi h_{11/2}^2\otimes \nu h_{11/2}^{-1}$ because of the crossing between the neutron levels $1h_{11/2}$ and $1h_{9/2}$. 
These results suggest that there is no chiral rotation up to $\hbar\omega=0.55$MeV, similarly as Ref. \cite{PengJ2020PLB}.

In order to investigate the pairing effects on the critical frequency, the evolution of the azimuth angle $\varphi$ are shown for different proton (left panel) and neutron (right panel) pairing strengths. 
In the left panel, when the proton pairing strength is enhanced by 50\%, 100\% or 150\%, the azimuth angle $\varphi$ respectively becomes nonzero at 0.52, 0.49 or 0.47 MeV, indicating the occurrence of the chiral rotation. 
These results suggest that the pairing correlations could induce the early appearance of the chiral rotation. 
Similarly, in the right panel, for the enhanced neutron pairing, the azimuth angle $\varphi$ respectively becomes nonzero at 0.54, 0.53 and 0.52 MeV. 
Obviously, the chiral critical frequency is more sensitive to the proton pairing than neutron for the 3-qp band in $^{135}$Nd. 

\begin{figure}[t]
  \centering
  \includegraphics[width=\linewidth]{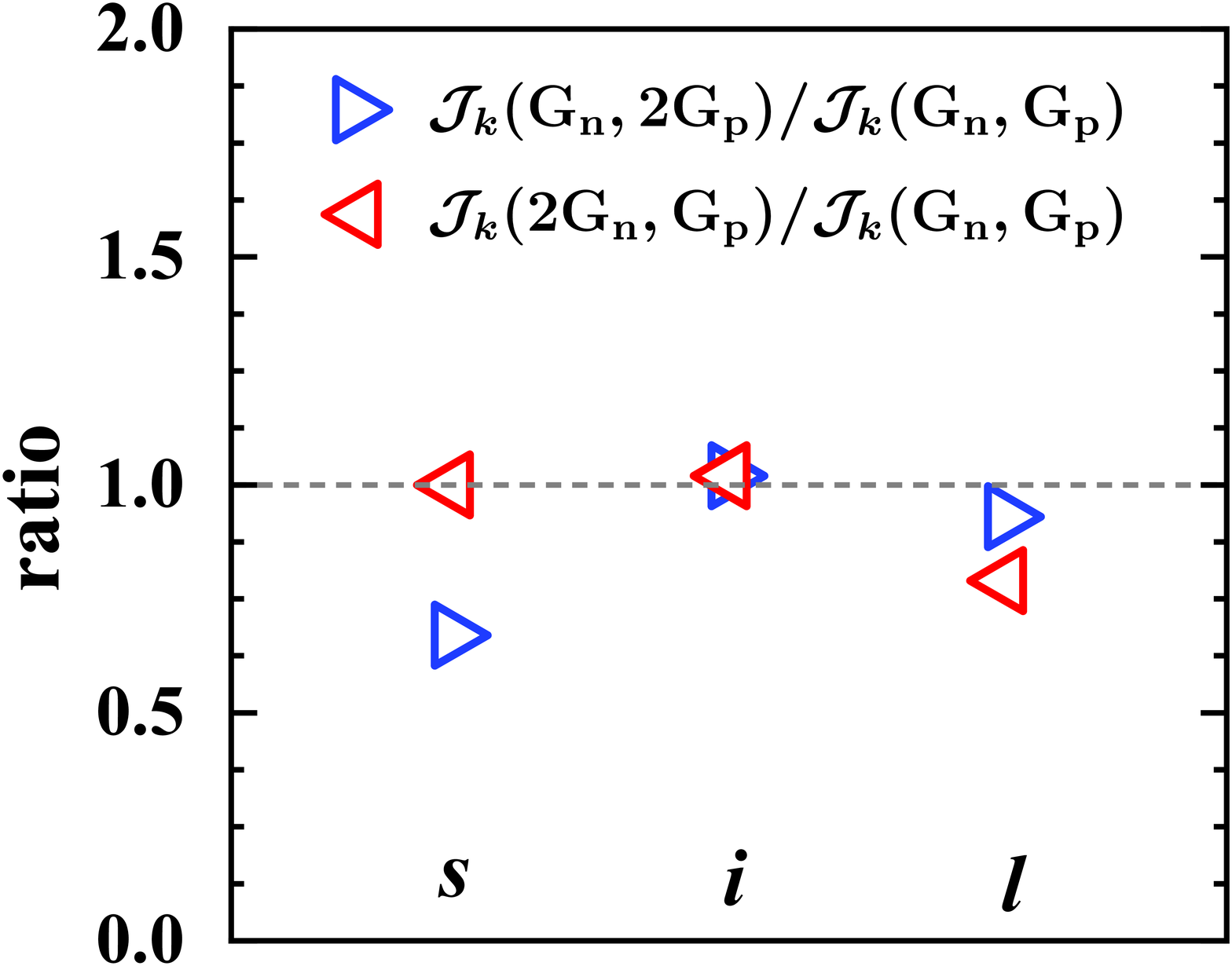}
  \caption{Ratios of the MOIs along the $s$, $i$ and $l$ axes $\mathcal{J}_k~(k=s,i,l)$  
           with the pairing strengths $(\mathrm{G}_\mathrm{n}, 2\mathrm{G}_\mathrm{p})$ 
           or $(2\mathrm{G}_\mathrm{n}, \mathrm{G}_\mathrm{p})$ relative to 
           those with the pairing strengths $(\mathrm{G}_\mathrm{n},\mathrm{G}_\mathrm{p})$.}
  \label{fig3}
\end{figure}

In order to understand the mechanism of the pairing effects on the critical frequency, the moments of inertia (MoIs) along the three principal axes will be analyzed.  

The MoIs along the three principal axes, $\mathcal{J}_k$, can be extracted from the corresponding angular momentum alignments $J_k$ and  rotational frequency $\omega_k$, $\mathcal{J}_k=dJ_k/d\omega_k, k = s, i, l$. The $J_k(\omega_k)$ relations can be obtained by cranking about the $s$, $i$, $l$ axes, respectively. In practice, it’s difficult to achieve convergent results for the configuration $\pi h_{11/2}^2\otimes \nu h_{11/2}^{-1}$ along the $i-$ and $l-$ axes. 
The calculations are performed for $\boldsymbol{\omega}$ at $(90^{\circ}, 20^{\circ})$, and the components $J_i$ and $\omega_i$ are used to calculate $\mathcal{J}_i$. 
Similarly, $\mathcal{J}_l$ is obtained from the calculation for $\boldsymbol{\omega}$ at $(70^{\circ}, 0^{\circ})$. 
The relations of the obtained $J_k$ and $\omega_k$ are approximately linear, and the corresponding slopes are taken as $\mathcal{J}_k$.
When the pairing strengths are $(\text{G}_\text{n}, \text{G}_\text{p})$, $\mathcal{J}_s = 17.2\hbar^2/$MeV, $\mathcal{J}_i = 36.1\hbar^2/$MeV and $\mathcal{J}_l = 18.7\hbar^2/$MeV.  

The ratios of the MoIs along the $s$, $i$ and $l$ axes with the pairing strengths $(\mathrm{G}_\mathrm{n}, 2\mathrm{G}_\mathrm{p})$ or $(2\mathrm{G}_\mathrm{n}, \mathrm{G}_\mathrm{p})$, relative to those when the pairing strengths are $(\text{G}_\text{n}, \text{G}_\text{p})$, are presented in Fig. \ref{fig3}. 
Compared with the MOIs with the pairing strengths $(\text{G}_\text{n}, \text{G}_\text{p})$, for $(\mathrm{G}_\mathrm{n}, 2\mathrm{G}_\mathrm{p})$, the $\mathcal{J}_s$ is significantly suppressed, by 33$\%$, and the $\mathcal{J}_i$ and $\mathcal{J}_l$ are almost unchanged. 
For $(2\mathrm{G}_\mathrm{n}, \mathrm{G}_\mathrm{p})$, the $\mathcal{J}_l$ is suppressed by 20$\%$, and the $\mathcal{J}_s$ and $\mathcal{J}_i$ are almost unchanged. 
The suppression of $\mathcal{J}_s$ or $\mathcal{J}_l$ results in the enhanced preference of the collective rotation along the $i$ axis, and induces the early appearance of the chiral rotation. 
It is shown in Fig. \ref{fig2} that the critical frequency is more sensitive to the proton pairing than neutron. 
This can be understood from the 33\% suppression of $\mathcal{J}_s$ by the proton pairing and the 20\% suppression of $\mathcal{J}_l$ by the neutron pairing in Fig. \ref{fig3}.

\begin{figure}[t]
  \centering
  \includegraphics[width=\linewidth]{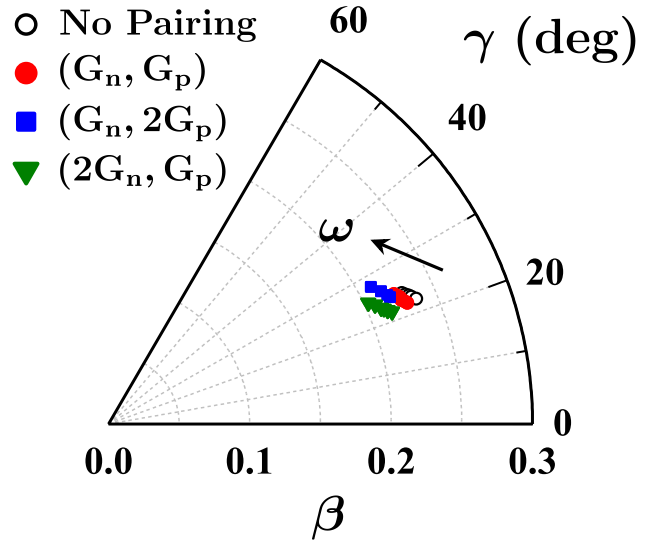}
  \caption{Evolution of the deformation parameters $\beta$ and $\gamma$ with the rotational frequency $\omega$ 
           for the pairing strengths $(\text{G}_\text{n}, \text{G}_\text{p})$, 
           $(\text{G}_\text{n}, 2\text{G}_\text{p})$, 
           and $(2\text{G}_\text{n}, \text{G}_\text{p})$, in comparison with the results without pairing.}
  \label{fig4}
\end{figure}

The evolution of the deformation parameters $\beta$ and $\gamma$ with the rotational frequency is shown in Fig. \ref{fig4}, for the pairing strengths $(\text{G}_\text{n}, \text{G}_\text{p})$, $(\text{G}_\text{n}, 2\text{G}_\text{p})$, and $(2\text{G}_\text{n}, \text{G}_\text{p})$, in comparison with the results without pairing. 
The influence of the pairing strengths on the deformation parameters is negligible, i.e., within $\Delta\beta=0.04$, and $\Delta\gamma=6^{\circ}$. 
Accordingly, the MOIs for the collective rotation remain unchanged. Since $\mathcal{J}_i$ mainly comes from the contribution of the collective rotation, its independence on the pairing strength in Fig. \ref{fig3} can be understood. 
For the same reason, the changes in $\mathcal{J}_s$ and $\mathcal{J}_l$ do not come from the collective rotation. 
For the 3-qp band, the angular momentum along the $s$ axis is mainly contributed from the proton particles, and the angular momentum along the $l$ axis is mainly contributed from the neutron hole. 
Therefore, the suppression of $\mathcal{J}_s$/$\mathcal{J}_l$ is due to the reduction of the particle/hole alignments along the $s$/$l$ axis by the proton/neutron pairing.

\section{Summary}

In summary, based on the 3D cranking CDFT, the SLAP with exact particle number conservation is implemented to take into account the pairing correlations and applied for the chiral doublet bands built on the configuration $\pi h_{11/2}^2\otimes \nu h_{11/2}^{-1}$ in $^{135}$Nd. 
The longstanding and challenging problem about the microscopic understanding on the influence of the pairing correlations on the nuclear chiral rotation is addressed.

The data available, including the $I-\omega$ relation, as well as the electromagnetic transition probabilities $B(M1)$ and $B(E2)$, are well reproduced. 
It is found that the superfluidity can reduce the critical frequency and make the chiral rotation easier. 
The mechanism is analyzed by investigating the pairing effects on the MOIs along the three principal axes and the deformation parameters $\beta$ and $\gamma$. 
It is demonstrated that the particle/hole alignments along the $s$/$l$ axis are reduced by the pairing correlations, resulting in the enhanced preference of the collective rotation along the $i$ axis, and inducing the early appearance of the chiral rotation.


\section*{Acknowledgments}

Y. P. Wang is grateful to B. W. Xiong for his help in the numerical implementation. Fruitful discussions with Y. K. Wang, S. Q. Zhang, and P. W. Zhao are acknowledged. This work was partly supported by the National Key R\&D Program of China (Contract No. 2018YFA0404400), the National Natural Science Foundation of China (Grants No. 11875075, No. 11935003, No. 11975031, No. 12141501, No. 12070131001), and the High-performance Computing Platform of Peking University.


\section*{References}

\begin{thebibliography}{10}
\expandafter\ifx\csname url\endcsname\relax
  \def\url#1{\texttt{#1}}\fi
\expandafter\ifx\csname urlprefix\endcsname\relax\def\urlprefix{URL }\fi
\expandafter\ifx\csname href\endcsname\relax
  \def\href#1#2{#2} \def\path#1{#1}\fi

\bibitem{Frauendorf1997NPA}
  S.~Frauendorf, J.~Meng, {Tilted rotation of triaxial nuclei}, Nucl. Phys. A
  617~(2) (1997) 131--147.
\newblock \href {https://doi.org/10.1016/S0375-9474(97)00004-3}
  {\path{doi:10.1016/S0375-9474(97)00004-3}}.

\bibitem{Starosta2001PRL}
  K.~Starosta, T.~Koike, C.~J. Chiara, D.~B. Fossan, D.~R. LaFosse, A.~A. Hecht,
  C.~W. Beausang, M.~A. Caprio, J.~R. Cooper, R.~Krücken, J.~R. Novak, N.~V.
  Zamfir, K.~E. Zyromski, D.~J. Hartley, D.~L. Balabanski, J.~Y. Zhang,
  S.~Frauendorf, V.~I. Dimitrov,
  {{Chiral
  Doublet Structures in Odd-Odd N = 75 Isotones: Chiral Vibrations}}, Phys.
  Rev. Lett. 86~(6) (2001) 971--974.
\newblock \href {https://doi.org/10.1103/PhysRevLett.86.971}
  {\path{doi:10.1103/PhysRevLett.86.971}}.

\bibitem{Tonev2006PRL}
  D.~Tonev, G.~de~Angelis, P.~Petkov, A.~Dewald, S.~Brant, S.~Frauendorf, D.~L.
  Balabanski, P.~Pejovic, D.~Bazzacco, P.~Bednarczyk, F.~Camera, A.~Fitzler,
  A.~Gadea, S.~Lenzi, S.~Lunardi, N.~Marginean, O.~M\"oller, D.~R. Napoli,
  A.~Paleni, C.~M. Petrache, G.~Prete, K.~O. Zell, Y.~H. Zhang, J.~Y. Zhang,
  Q.~Zhong, D.~Curien,
  {{Transition
  Probabilities in $^{134}\mathrm{Pr}$: A Test for Chirality in Nuclear
  Systems}}, Phys. Rev. Lett. 96 (2006) 052501.
\newblock \href {https://doi.org/10.1103/PhysRevLett.96.052501}
  {\path{doi:10.1103/PhysRevLett.96.052501}}.

\bibitem{Petrache2006PRL}
  C.~M. Petrache, G.~B. Hagemann, I.~Hamamoto, K.~Starosta,
  {{Risk of
  Misinterpretation of Nearly Degenerate Pair Bands as Chiral Partners in
  Nuclei}}, Phys. Rev. Lett. 96 (2006) 112502.
\newblock \href {https://doi.org/10.1103/PhysRevLett.96.112502}
  {\path{doi:10.1103/PhysRevLett.96.112502}}.

\bibitem{MengJ2006PRC}
  J.~Meng, J.~Peng, S.~Q. Zhang, S.~G. Zhou, {Possible existence of multiple
  chiral doublets in $^{106}\mathrm{R}$h}, Phys. Rev. C 73~(3) (2006) 037303.
\newblock \href {https://doi.org/10.1103/PhysRevC.73.037303}
  {\path{doi:10.1103/PhysRevC.73.037303}}.

\bibitem{Ayangeakaa2013PRL}
  A.~D. Ayangeakaa, U.~Garg, M.~D. Anthony, S.~Frauendorf, J.~T. Matta, B.~K.
  Nayak, D.~Patel, Q.~B. Chen, S.~Q. Zhang, P.~W. Zhao, B.~Qi, J.~Meng,
  R.~V.~F. Janssens, M.~P. Carpenter, C.~J. Chiara, F.~G. Kondev, T.~Lauritsen,
  D.~Seweryniak, S.~Zhu, S.~S. Ghugre, R.~Palit, {Evidence for Multiple Chiral
  Doublet Bands in $^{133}\mathrm{C}$e}, Phys. Rev. Lett. 110~(17) (2013)
  172504.
\newblock \href {https://doi.org/10.1103/PHYSREVLETT.110.172504}
  {\path{doi:10.1103/PHYSREVLETT.110.172504}}.

\bibitem{Lieder2014PRL}
  E.~O. Lieder, R.~M. Lieder, R.~A. Bark, Q.~B. Chen, S.~Q. Zhang, J.~Meng, E.~A.
  Lawrie, J.~J. Lawrie, S.~P. Bvumbi, N.~Y. Kheswa, S.~S. Ntshangase, T.~E.
  Madiba, P.~L. Masiteng, S.~M. Mullins, S.~Murray, P.~Papka, D.~G. Roux,
  O.~Shirinda, Z.~H. Zhang, P.~W. Zhao, Z.~P. Li, J.~Peng, B.~Qi, S.~Y. Wang,
  Z.~G. Xiao, C.~Xu,
  {{Resolution of
  Chiral Conundrum in $^{106}\mathrm{Ag}$: Doppler-Shift Lifetime
  Investigation}}, Phys. Rev. Lett. 112 (2014) 202502.
\newblock \href {https://doi.org/10.1103/PhysRevLett.112.202502}
  {\path{doi:10.1103/PhysRevLett.112.202502}}.

\bibitem{LiuC2016PRL}
  C.~Liu, S.~Y. Wang, R.~A. Bark, S.~Q. Zhang, J.~Meng, B.~Qi, P.~Jones, S.~M.
  Wyngaardt, J.~Zhao, C.~Xu, S.~G. Zhou, S.~Wang, D.~P. Sun, L.~Liu, Z.~Q. Li,
  N.~B. Zhang, H.~Jia, X.~Q. Li, H.~Hua, Q.~B. Chen, Z.~G. Xiao, H.~J. Li,
  L.~H. Zhu, T.~D. Bucher, T.~Dinoko, J.~Easton, K.~Juh\'asz, A.~Kamblawe,
  E.~Khaleel, N.~Khumalo, E.~A. Lawrie, J.~J. Lawrie, S.~N.~T. Majola, S.~M.
  Mullins, S.~Murray, J.~Ndayishimye, D.~Negi, S.~P. Noncolela, S.~S.
  Ntshangase, B.~M. Nyak\'o, J.~N. Orce, P.~Papka, J.~F. Sharpey-Schafer,
  O.~Shirinda, P.~Sithole, M.~A. Stankiewicz, M.~Wiedeking, {Evidence for
  Octupole Correlations in Multiple Chiral Doublet Bands}, Phys. Rev. Lett. 116
  (2016) 112501.
\newblock \href {https://doi.org/10.1103/PhysRevLett.116.112501}
  {\path{doi:10.1103/PhysRevLett.116.112501}}.

\bibitem{XiongBW2019ADNDT}
  B.~W. Xiong, Y.~Y. Wang, {Nuclear chiral doublet bands data tables}, Atom. Data
  Nucl. Data Tabl. 125 (2019) 193--225.
\newblock \href {https://doi.org/10.1016/j.adt.2018.05.002}
  {\path{doi:10.1016/j.adt.2018.05.002}}.

\bibitem{PengJ2003PRC}
  J.~Peng, J.~Meng, S.~Q. Zhang, {Description of chiral doublets in
  $\mathrm{A}\sim$130 nuclei and the possible chiral doublets in
  $\mathrm{A}\sim$100 nuclei}, Phys. Rev. C 68~(4) (2003) 044324.
\newblock \href {https://doi.org/10.1103/PhysRevC.68.044324}
  {\path{doi:10.1103/PhysRevC.68.044324}}.

\bibitem{KoikeT2004PRL}
  T.~Koike, K.~Starosta, I.~Hamamoto, {Chiral Bands, Dynamical Spontaneous
  Symmetry Breaking, and the Selection Rule for Electromagnetic Transitions in
  the Chiral Geometry}, Phys. Rev. Lett. 93~(17) (2004) 172502.
\newblock \href {https://doi.org/10.1103/PhysRevLett.93.172502}
  {\path{doi:10.1103/PhysRevLett.93.172502}}.

\bibitem{ZhangSQ2007PRC}
  S.~Q. Zhang, B.~Qi, S.~Y. Wang, J.~Meng, {Chiral bands for a quasi-proton and
  quasi-neutron coupled with a triaxial rotor}, Phys. Rev. C 75 (2007) 044307.
\newblock \href {https://doi.org/10.1103/PhysRevC.75.044307}
  {\path{doi:10.1103/PhysRevC.75.044307}}.

\bibitem{QiB2009PLB}
  B.~Qi, S.~Q. Zhang, J.~Meng, S.~Y. Wang, S.~Frauendorf, {Chirality in odd-A
  nucleus $^{135}\mathrm{N}$d in particle rotor model}, Phys. Lett. B 675
  (2009) 175 -- 180.
\newblock \href {https://doi.org/10.1016/j.physletb.2009.02.061}
  {\path{doi:10.1016/j.physletb.2009.02.061}}.

\bibitem{ChenQB2018PLB}
  Q.~B. Chen, B.~F. Lv, C.~M. Petrache, J.~Meng, {Multiple chiral doublets in
  four-$j$ shells particle rotor model: Five possible chiral doublets in
  $^{136}_{~60}$Nd$_{76}$}, Phys. Lett. B 782 (2018) 744--749.
\newblock \href {https://doi.org/10.1016/j.physletb.2018.06.030}
  {\path{doi:10.1016/j.physletb.2018.06.030}}.

\bibitem{WangYY2019PLB}
  Y.~Y. Wang, S.~Q. Zhang, P.~W. Zhao, J.~Meng,
  {{Multiple
  chiral doublet bands with octupole correlations in reflection-asymmetric
  triaxial particle rotor model}}, Phys. Lett. B 792 (2019) 454 -- 460.
\newblock \href
  {https://doi.org/https://doi.org/10.1016/j.physletb.2019.04.014}
  {\path{doi:https://doi.org/10.1016/j.physletb.2019.04.014}}.

\bibitem{WangYP2020PRC}
  Y.~P. Wang, Y.~Y. Wang, J.~Meng,
  {{Pseudospin
  symmetry and octupole correlations for multiple chiral doublets in
  $^{131}\mathrm{Ba}$}}, Phys. Rev. C 102 (2020) 024313.
\newblock \href {https://doi.org/10.1103/PhysRevC.102.024313}
  {\path{doi:10.1103/PhysRevC.102.024313}}.

\bibitem{Madokoro2000PRC}
  H.~Madokoro, J.~Meng, M.~Matsuzaki, S.~Yamaji,
  {{Relativistic mean
  field description for the shears band mechanism in ${}^{84}\mathrm{Rb}$}},
  Phys. Rev. C 62 (2000) 061301(R).
\newblock \href {https://doi.org/10.1103/PhysRevC.62.061301}
  {\path{doi:10.1103/PhysRevC.62.061301}}.

\bibitem{Dimitrov2000PRL}
  V.~I. Dimitrov, S.~Frauendorf, F.~D$\mathrm{\ddot{o}}$nau, {Chirality of
  Nuclear Rotation}, Phys. Rev. Lett. 84~(25) (2000) 5732--5735.
\newblock \href {https://doi.org/10.1103/PhysRevLett.84.5732}
  {\path{doi:10.1103/PhysRevLett.84.5732}}.

\bibitem{Olbratowski2004PRL}
  P.~Olbratowski, J.~Dobaczewski, J.~Dudek, W.~P{\l}$\mathrm{\acute{o}}$ciennik,
  {Critical Frequency in Nuclear Chiral Rotation}, Phys. Rev. Lett. 93~(5)
  (2004) 052501.
\newblock \href {https://doi.org/10.1103/PhysRevLett.93.052501}
  {\path{doi:10.1103/PhysRevLett.93.052501}}.

\bibitem{ZhaoPW2017PLB}
  P.~W. Zhao, {Multiple chirality in nuclear rotation: a microscopic view}, Phys.
  Lett. B 773 (2017) 1.
\newblock \href {https://doi.org/10.1016/j.physletb.2017.08.001}
  {\path{doi:10.1016/j.physletb.2017.08.001}}.

\bibitem{PengJ2020PLB}
  J.~Peng, Q.~B. Chen, {Covariant density functional theory for nuclear chirality
  in $^{135}\mathrm{Nd}$}, Phys. Lett. B 810 (2020) 135795.
\newblock \href {https://doi.org/10.1016/j.physletb.2020.135795}
  {\path{doi:10.1016/j.physletb.2020.135795}}.

\bibitem{Mukhopadhyay2007PRL}
  S.~Mukhopadhyay, D.~Almehed, U.~Garg, S.~Frauendorf, T.~Li, P.~V.~M. Rao,
  X.~Wang, S.~S. Ghugre, M.~P. Carpenter, S.~Gros, A.~Hecht, R.~V.~F. Janssens,
  F.~G. Kondev, T.~Lauritsen, D.~Seweryniak, S.~Zhu,
  {{From Chiral
  Vibration to Static Chirality in $^{135}\mathrm{Nd}$}}, Phys. Rev. Lett. 99
  (2007) 172501.
\newblock \href {https://doi.org/10.1103/PhysRevLett.99.172501}
  {\path{doi:10.1103/PhysRevLett.99.172501}}.

\bibitem{Almehed2011PRC}
  D.~Almehed, F.~D$\mathrm{\ddot{o}}$nau, S.~Frauendorf, {Chiral vibrations in
  the $\mathit{A}$ = 135 region}, Phys. Rev. C 83~(5) (2011) 054308.
\newblock \href {https://doi.org/10.1103/PhysRevC.83.054308}
  {\path{doi:10.1103/PhysRevC.83.054308}}.

\bibitem{ChenQB2013PRC}
  Q.~B. Chen, S.~Q. Zhang, P.~W. Zhao, R.~V. Jolos, J.~Meng, {Collective
  Hamiltonian for chiral modes}, Phys. Rev. C 87~(2) (2013) 024314.
\newblock \href {https://doi.org/10.1103/PhysRevC.87.024314}
  {\path{doi:10.1103/PhysRevC.87.024314}}.

\bibitem{ChenQB2016PRC}
  Q.~B. Chen, S.~Q. Zhang, P.~W. Zhao, R.~V. Jolos, J.~Meng, {Two-dimensional
  collective Hamiltonian for chiral and wobbling modes}, Phys. Rev. C 94 (2016)
  044301.
\newblock \href {https://doi.org/10.1103/PhysRevC.94.044301}
  {\path{doi:10.1103/PhysRevC.94.044301}}.

\bibitem{ChenQB2018PRC}
  Q.~B. Chen, J.~Meng, {Reexamine the nuclear chiral geometry from the
  orientation of the angular momentum}, Phys. Rev. C 98~(3) (2018) 031303(R).
\newblock \href {https://doi.org/10.1103/PhysRevC.98.031303}
  {\path{doi:10.1103/PhysRevC.98.031303}}.

\bibitem{Brant2008PRC}
  S.~Brant, D.~Tonev, G.~de~Angelis, A.~Ventura, {Dynamic chirality in the
  interacting boson fermion-fermion model}, Phys. Rev. C 78 (2008) 034301.
\newblock \href {https://doi.org/10.1103/PhysRevC.78.034301}
  {\path{doi:10.1103/PhysRevC.78.034301}}.

\bibitem{Raduta2016JPG}
  A.~A. Raduta, A.~H. Raduta, C.~M. Petrache, {New type of chiral motion in
  even-even nuclei: the $^{138}$Nd case}, J. Phys. G: Nucl. Part. Phys. 43~(9)
  (2016) 095107.
\newblock \href {https://doi.org/10.1088/0954-3899/43/9/095107}
  {\path{doi:10.1088/0954-3899/43/9/095107}}.

\bibitem{Hara1995IJMPE}
  K.~Hara, Y.~Sun, {Projected Shell Model and High-Spin Spectroscopy}, Int. J.
  Mod. Phys. E 4~(04) (1995) 637--785.
\newblock \href {https://doi.org/10.1142/S0218301395000250}
  {\path{doi:10.1142/S0218301395000250}}.

\bibitem{Bhat2014NPA}
  G.~H. Bhat, R.~N. Ali, J.~A. Sheikh, R.~Palit, {Investigation of doublet-bands
  in $^{124,126,130,132}\mathrm{C}$s odd-odd nuclei using triaxial projected
  shell model approach}, Nucl. Phys. A 922 (2014) 150--162.
\newblock \href {https://doi.org/10.1016/j.nuclphysa.2013.12.006}
  {\path{doi:10.1016/j.nuclphysa.2013.12.006}}.

\bibitem{ChenFQ2017PRC}
  F.~Q. Chen, Q.~B. Chen, Y.~A. Luo, J.~Meng, S.~Q. Zhang,
  {{Chiral geometry
  in symmetry-restored states: Chiral doublet bands in ${}^{128}\mathrm{C}$s}},
  Phys. Rev. C 96 (2017) 051303(R).
\newblock \href {https://doi.org/10.1103/PhysRevC.96.051303}
  {\path{doi:10.1103/PhysRevC.96.051303}}.

\bibitem{ChenFQ2018PLB}
  F.~Q. Chen, J.~Meng, S.~Q. Zhang, {Chiral geometry and rotational structure for
  $^{130}$Cs in the projected shell model}, Phys. Lett. B 785 (2018) 211--216.
\newblock \href {https://doi.org/10.1016/j.physletb.2018.08.039}
  {\path{doi:10.1016/j.physletb.2018.08.039}}.

\bibitem{WangYK2019PRC}
  Y.~K. Wang, F.~Q. Chen, P.~W. Zhao, S.~Q. Zhang, J.~Meng,
  {{Multichiral
  facets in symmetry restored states: Five chiral doublet candidates in the
  even-even nucleus $^{136}\mathrm{Nd}$}}, Phys. Rev. C 99 (2019) 054303.
\newblock \href {https://doi.org/10.1103/PhysRevC.99.054303}
  {\path{doi:10.1103/PhysRevC.99.054303}}.

\bibitem{MengJ2013FP}
  J.~Meng, J.~Peng, S.~Q. Zhang, P.~W. Zhao,
  {{Progress on tilted axis
  cranking covariant density functional theory for nuclear magnetic and
  antimagnetic rotation}}, Front. Phys. 8 (2013) 55--79.
\newblock \href {https://doi.org/10.1007/s11467-013-0287-y}
  {\path{doi:10.1007/s11467-013-0287-y}}.

\bibitem{ZhaoPW2011PRL}
  P.~W. Zhao, J.~Peng, H.~Z. Liang, P.~Ring, J.~Meng, {Antimagnetic Rotation Band
  in Nuclei: A Microscopic Description}, Phys. Rev. Lett. 107 (2011) 122501.
\newblock \href {https://doi.org/10.1103/PhysRevLett.107.122501}
  {\path{doi:10.1103/PhysRevLett.107.122501}}.

\bibitem{Olbratowski2006PRC}
  P.~Olbratowski, J.~Dobaczewski, J.~Dudek, {Search for the Skyrme-Hartree-Fock
  Solutions for Chiral Rotation in $\mathit{N}$ = 75 Isotones}, Phys. Rev. C
  73~(5) (2006) 054308.
\newblock \href {https://doi.org/10.1103/PhysRevC.73.054308}
  {\path{doi:10.1103/PhysRevC.73.054308}}.

\bibitem{ZhaoPW2015PRL}
  P.~W. Zhao, N.~Itagaki, J.~Meng,
  {{Rod-shaped
  Nuclei at Extreme Spin and Isospin}}, Phys. Rev. Lett. 115 (2015) 022501.
\newblock \href {https://doi.org/10.1103/PhysRevLett.115.022501}
  {\path{doi:10.1103/PhysRevLett.115.022501}}.

\bibitem{Ring1996PPNP}
  P.~Ring,
  {{Relativistic
  mean field theory in finite nuclei}}, Progress in Particle and Nuclear
  Physics 37 (1996) 193--263.
\newblock \href {https://doi.org/https://doi.org/10.1016/0146-6410(96)00054-3}
  {\path{doi:https://doi.org/10.1016/0146-6410(96)00054-3}}.

\bibitem{Vretenar2005PR}
  D.~Vretenar, A.~Afanasjev, G.~Lalazissis, P.~Ring,
  {{Relativistic
  Hartree–Bogoliubov theory: static and dynamic aspects of exotic nuclear
  structure}}, Physics Reports 409~(3) (2005) 101--259.
\newblock \href {https://doi.org/https://doi.org/10.1016/j.physrep.2004.10.001}
  {\path{doi:https://doi.org/10.1016/j.physrep.2004.10.001}}.

\bibitem{MengJ2006PPNP}
  J.~Meng, H.~Toki, S.~G. Zhou, S.~Q. Zhang, W.~H. Long, L.~S. Geng,
  {{Relativistic
  continuum Hartree Bogoliubov theory for ground-state properties of exotic
  nuclei}}, Progress in Particle and Nuclear Physics 57~(2) (2006) 470--563.
\newblock \href {https://doi.org/https://doi.org/10.1016/j.ppnp.2005.06.001}
  {\path{doi:https://doi.org/10.1016/j.ppnp.2005.06.001}}.

\bibitem{Niksic2011PPNP}
  T.~Nikšić, D.~Vretenar, P.~Ring,
  {{Relativistic
  nuclear energy density functionals: Mean-field and beyond}}, Progress in
  Particle and Nuclear Physics 66~(3) (2011) 519--548.
\newblock \href {https://doi.org/https://doi.org/10.1016/j.ppnp.2011.01.055}
  {\path{doi:https://doi.org/10.1016/j.ppnp.2011.01.055}}.

\bibitem{MengJ2016Relativistic}
  J.~Meng (Ed.), {Relativistic Density Functional for Nuclear Structure}, Vol.~10
  of International Review of Nuclear Physics, World Scientific, Singapore,
  2016.

\bibitem{PengJ2008PRC2}
  J.~Peng, J.~Meng, P.~Ring, S.~Q. Zhang,
  {{Covariant density
  functional theory for magnetic rotation}}, Phys. Rev. C 78 (2008) 024313.
\newblock \href {https://doi.org/10.1103/PhysRevC.78.024313}
  {\path{doi:10.1103/PhysRevC.78.024313}}.

\bibitem{ZengJY1983NPA}
  J.~Y. Zeng, T.~S. Cheng,
  {{Particle-number-conserving
  method for treating the nuclear pairing correlation}}, Nucl. Phys. A 405~(1)
  (1983) 1--28.
\newblock \href {https://doi.org/https://doi.org/10.1016/0375-9474(83)90320-2}
  {\path{doi:https://doi.org/10.1016/0375-9474(83)90320-2}}.

\bibitem{ZengJY1994PRC}
  J.~Y. Zeng, T.~H. Jin, Z.~J. Zhao,
  {{Reduction of
  nuclear moment of inertia due to pairing interaction}}, Phys. Rev. C 50
  (1994) 1388--1397.
\newblock \href {https://doi.org/10.1103/PhysRevC.50.1388}
  {\path{doi:10.1103/PhysRevC.50.1388}}.

\bibitem{MengJ2006FPC}
  J.~Meng, J.~Y. Guo, L.~Liu, S.~Q. Zhang,
  {{Shell-model-like
  Approach (SLAP) for the Nuclear Properties in Relativistic Mean Field
  Theory}}, Front. Phys. China 1 (2006) 38--46.
\newblock \href {https://doi.org/10.1007/s11467-005-0013-5}
  {\path{doi:10.1007/s11467-005-0013-5}}.

\bibitem{ShiZ2018PRC}
  Z.~Shi, Z.~H. Zhang, Q.~B. Chen, S.~Q. Zhang, J.~Meng,
  {{Shell-model-like
  approach based on cranking covariant density functional theory: Band crossing
  and shape evolution in $^{60}\mathrm{Fe}$}}, Phys. Rev. C 97 (2018) 034317.
\newblock \href {https://doi.org/10.1103/PhysRevC.97.034317}
  {\path{doi:10.1103/PhysRevC.97.034317}}.

\bibitem{ZhaoPW2010PRC}
  P.~W. Zhao, Z.~P. Li, J.~M. Yao, J.~Meng, {New parametrization for the nuclear
  covariant energy density functional with a point-coupling interaction}, Phys.
  Rev. C 82 (2010) 054319.
\newblock \href {https://doi.org/10.1103/PhysRevC.82.054319}
  {\path{doi:10.1103/PhysRevC.82.054319}}.

\bibitem{WangYK2017PRC}
  Y.~K. Wang, {{Yrast
  band of $^{109}\mathrm{Ag}$ described by tilted axis cranking covariant
  density functional theory with a separable pairing force}}, Phys. Rev. C 96
  (2017) 054324.
\newblock \href {https://doi.org/10.1103/PhysRevC.96.054324}
  {\path{doi:10.1103/PhysRevC.96.054324}}.

\bibitem{ZhuS2003PRL}
  S.~Zhu, U.~Garg, B.~K. Nayak, S.~S. Ghugre, N.~S. Pattabiraman, D.~B. Fossan,
  T.~Koike, K.~Starosta, C.~Vaman, R.~V.~F. Janssens, R.~S. Chakrawarthy,
  M.~Whitehead, A.~O. Macchiavelli, S.~Frauendorf,
  {{A Composite
  Chiral Pair of Rotational Bands in the Odd-$A$ Nucleus
  $^{135}\mathrm{N}\mathrm{d}$}}, Phys. Rev. Lett. 91 (2003) 132501.
\newblock \href {https://doi.org/10.1103/PhysRevLett.91.132501}
  {\path{doi:10.1103/PhysRevLett.91.132501}}.

\bibitem{LvBF2019PRC}
  B.~F. Lv, C.~M. Petrache, Q.~B. Chen, J.~Meng, A.~Astier, E.~Dupont,
  P.~Greenlees, H.~Badran, T.~Calverley, D.~M. Cox, T.~Grahn, J.~Hilton,
  R.~Julin, S.~Juutinen, J.~Konki, J.~Pakarinen, P.~Papadakis, J.~Partanen,
  P.~Rahkila, P.~Ruotsalainen, M.~Sandzelius, J.~Saren, C.~Scholey, J.~Sorri,
  S.~Stolze, J.~Uusitalo, B.~Cederwall, A.~Ertoprak, H.~Liu, S.~Guo, M.~L. Liu,
  J.~G. Wang, X.~H. Zhou, I.~Kuti, J.~Timár, A.~Tucholski, J.~Srebrny,
  C.~Andreoiu, {Chirality of $^{135}\mathrm{Nd}$ reexamined: Evidence for
  multiple chiral doublet bands}, Phys. Rev. C 100 (2019) 024314.
\newblock \href {https://doi.org/10.1103/PhysRevC.100.024314}
  {\path{doi:10.1103/PhysRevC.100.024314}}.

\bibitem{Brant2009PRC}
  S.~Brant, C.~M. Petrache,
  {{Chiral bands in
  $^{135}\mathrm{Nd}$: The interacting boson-fermion model approach}}, Phys.
  Rev. C 79 (2009) 054326.
\newblock \href {https://doi.org/10.1103/PhysRevC.79.054326}
  {\path{doi:10.1103/PhysRevC.79.054326}}.

\bibitem{ZhaoPW2015PRC}
  P.~W. Zhao, S.~Q. Zhang, J.~Meng,
  {{Impact of pairing
  correlations on the orientation of the nuclear spin}}, Phys. Rev. C 92 (2015)
  034319.
\newblock \href {https://doi.org/10.1103/PhysRevC.92.034319}
  {\path{doi:10.1103/PhysRevC.92.034319}}.

\bibitem{RenZX2022PRC}
  Z.~X. Ren, P.~W. Zhao, J.~Meng,
  {{Dynamics of
  rotation in chiral nuclei}}, Phys. Rev. C 105 (2022) L011301.
\newblock \href {https://doi.org/10.1103/PhysRevC.105.L011301}
  {\path{doi:10.1103/PhysRevC.105.L011301}}.

\bibitem{Kerman1981NPA}
  A.~K. Kerman, N.~Onishi, 
  {{Nuclear rotations
  studied by the time-dependent variational method}}, Nucl. Phys. A 361 (1981) 179-191.
\newblock \href {https://doi.org/10.1016/0375-9474(81)90475-9}
  {\path{doi:10.1016/0375-9474(81)90475-9}}.

\bibitem{ShiY2013PRC}
  Y.~Shi, C.~L. Zhang, J.~Dobaczewski, W.~Nazarewicz, 
  {{Kerman-Onishi conditions
  in self-consistent tilted-axis-cranking mean-field calculations}}, Phys. Rev. C 88 (2013) 034311.
\newblock \href {https://link.aps.org/doi/10.1103/PhysRevC.88.034311}
  {\path{doi:10.1103/PhysRevC.88.034311}}.

\bibitem{ShiY2012PRL}
  Y.~Shi, J.~Dobaczewski, S.~Frauendorf, W.~Nazarewicz, J.~C.~Pei, F.~R.~Xu, N.~Nikolov, 
  {{Self-Consistent Tilted-Axis-Cranking
  Study of Triaxial Strongly Deformed Bands in $^{158}\mathrm{Er}$ at Ultrahigh Spin}}, Phys. Rev. Lett. 108 (2012) 092501.
\newblock \href {https://link.aps.org/doi/10.1103/PhysRevLett.108.092501}
  {\path{doi:10.1103/PhysRevLett.108.092501}}.

\bibitem{Beck1987PRL}
  E.~M. Beck, F.~S. Stephens, J.~C. Bacelar, M.~A. Deleplanque, R.~M. Diamond,
  J.~E. Draper, C.~Duyar, R.~J. McDonald,
  {{Superdeformed
  Band in $^{135}\mathrm{Nd}$}}, Phys. Rev. Lett. 58 (1987) 2182--2185.
\newblock \href {https://doi.org/10.1103/PhysRevLett.58.2182}
  {\path{doi:10.1103/PhysRevLett.58.2182}}.

\end{thebibliography}
\end{document}